\documentclass[prd, twocolumn, superscriptaddress,floatfix, nofootinbib, preprintnumbers]{revtex4-2}
\usepackage[utf8]{inputenc}
\usepackage[colorlinks=true,citecolor=blue]{hyperref}
\usepackage{amssymb}
\usepackage{amsmath}
\usepackage{graphicx}
\usepackage{footmisc}
\usepackage{url}
\usepackage{multirow}
\usepackage{makecell}
\usepackage{enumerate}
\usepackage{hhline}
\usepackage{comment}
\usepackage{nicefrac}

\def\beq{\begin{equation}}
\def\eeq{\end{equation}}
\newcommand{\bea}{\begin{eqnarray}\begin{aligned}}
\newcommand{\eea}{\end{aligned}\end{eqnarray}}

\begin{document}

\title{
 Combining Resonant and Tail-based Anomaly Detection
}

\author{Gerrit Bickendorf}
\affiliation{Bethe Center for Theoretical Physics and Physikalisches
	Institut, Universit\"at Bonn, Nussallee~12, 53115 Bonn, Germany}

\author{Manuel Drees}
\affiliation{Bethe Center for Theoretical Physics and Physikalisches
	Institut, Universit\"at Bonn, Nussallee~12, 53115 Bonn, Germany}

\author{Gregor Kasieczka}
\affiliation{Institut f\"ur Experimentalphysik, Universit\"at Hamburg, Luruper Chaussee 149, 22761 Hamburg, Germany}

\author{Claudius Krause}
\affiliation{Institut f\"ur Theoretische Physik, Universit\"at Heidelberg, Philosophenweg 12, 69120 Heidelberg, Germany}

\author{David Shih}
\affiliation{New High Energy Theory Center, Rutgers University \\
  Piscataway, New Jersey 08854-8019, USA}

\begin{abstract}
In many well-motivated models of the electroweak scale, cascade decays of new particles can result in highly boosted hadronic resonances (e.g. $Z/W/h$). This can make these models rich and promising targets for recently developed resonant anomaly detection methods powered by modern machine learning. We demonstrate this using the state-of-the-art CATHODE method applied to supersymmetry scenarios with gluino pair production. We show that CATHODE, despite being model-agnostic, is nevertheless competitive with dedicated cut-based searches, while simultaneously covering a much wider region of parameter space. The gluino events also populate the tails of the missing energy and $H_T$ distributions, making this a novel combination of resonant and tail-based anomaly detection.

\end{abstract}

\maketitle

\section{Introduction}

The absence of new physics at the LHC is an enduring mystery. Many well-motivated theoretical frameworks such as supersymmetry, extra dimensions, and composite Higgs have predicted signatures of new particles at the weak scale, yet countless searches for these new particles have not found any significant evidence for them to date. 

Nearly all of these searches for physics beyond the Standard Model (BSM) are model-specific to some degree, optimized for specific signal scenarios, often using simulations. It is highly likely that these searches have not thoroughly covered the full phase space at the LHC, leaving a real possibility of new physics simply hiding in the data at the LHC, undiscovered because we have not searched for it.

Recently there has been considerable interest in developing more model-agnostic search strategies for the LHC~\cite{Karagiorgi:2021ngt,Kasieczka:2021xcg,Aarrestad:2021oeb}. In particular, a lot of activity has focused on ``resonant anomaly detection" methods~\cite{Collins:2018epr,Collins:2019jip,Nachman:2020lpy,Andreassen:2020nkr,Amram:2020ykb,ATLAS:2020iwa,Benkendorfer:2020gek,Stein:2020rou,Park:2020pak,Collins:2021nxn,Hallin:2021wme,Kamenik:2022qxs,Raine:2022hht,Kasieczka:2022naq,Hallin:2022eoq,Chen:2022suv,Golling:2022nkl,Golling:2023yjq,sengupta2023curtains}. 
In these approaches, one singles out a specific kinematic feature (e.g.\ the invariant mass of something in the event) in which new physics is postulated to be localized (resonant) to a window. This window serves as the signal region (SR) of the anomaly search. Then one uses the sidebands and modern machine learning techniques to learn a  multivariate, data-driven background template in additional features $x$. Finally, one employs further techniques (such as a classifier) to learn the difference between the background template and the data itself in the SR, in the form of an anomaly score 
\begin{equation}\label{eq:anomalyscore}
R(x) = \frac{p_{\rm data}(x)}{p_{\rm bg}(x)}
\end{equation}
If $p_{bg}(x)$ is the true background density and the classifier is optimal, this is the Neyman-Pearson optimal (idealized) anomaly detector in the SR. By cutting on $R(x)$, one can greatly enhance the significance of any resonant new physics in the SR. 

So far this activity has almost exclusively focused on new physics that is fully localized --- both in the SR and in the features $x$ --- and
using a global resonant feature such as the invariant mass of a dijet system.
Here we point out that the resonant anomaly detection technique is more general and both assumptions can be easily relaxed.

First, resonant anomaly detection methods can be applied to any resonant feature in the event, as long as the background satisfies the assumption of smoothness in that feature. One strong motivation for considering this broader perspective is that in many well-motivated models, such as those for the electroweak hierarchy, highly boosted resonances (either $Z/W/h$ from the SM or additional BSM particles such as a heavier Higgs boson)  can be quite common in the decays of heavier particles.

Additionally, we seek to broaden the scope of resonant anomaly detection in this work by pointing out that the signal need not be localized in all (or any) of the features $x$; it can also appear on the tails of the $x$ distribution (although of course the signal needs to be distinguishable in some of the features). This is a feature of resonant anomaly detection that has not been utilized so far. 
Anomalies on the tails of distributions such as $p_T^\text{miss}$, $H_T$ and $M_{eff}$ are quite common and plausible in models of TeV-scale new physics.

\begin{figure}[t]
\centering
\includegraphics[width=0.6\linewidth]{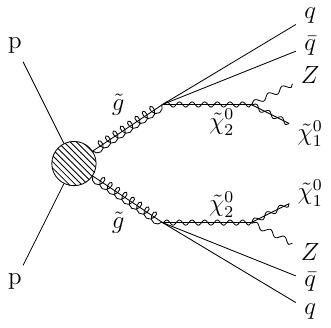}
\caption{Diagram of the signal process $p p \rightarrow \widetilde{g} \widetilde{g}$ with $ \widetilde{g}\rightarrow q \bar{q} \widetilde{\chi}_2^0, \widetilde{\chi}_2^0 \rightarrow Z \widetilde{\chi}_1^0$}
\label{fig:feynman_diag}
\end{figure}

In this work, we illustrate this broader application of resonant anomaly detection using a supersymmetric (SUSY) scenario as a well-motivated example. This SUSY scenario consists of gluino pair production, with the gluinos decaying to neutralinos plus a pair of (light quark) jets, and the neutralino decaying to another neutralino (LSP) through an on-shell $Z$ boson, as shown in figure \ref{fig:feynman_diag}. The LSP neutralino is much lighter than the second neutralino, meaning the $Z$'s are highly boosted. Therefore every event has two boosted $Z$'s, jets, and missing energy. CMS previously searched for this signal with a cut-based analysis~\cite{CMS:2020fia}. It defined a series of SRs requiring leading and subleading AK8 jets within $m\in[70,100]$~GeV and considering $p_T^\text{miss}$ in different exclusive bins. The background was estimated in two steps. First, the total number of events $B_{norm}$ in the SR was determined using sidebands in the leading AK8 jet mass (the subleading AK8 jet was required to be in the SR). Then the distribution in $p_T^\text{miss}$ bins (shape) is determined using the $p_T^\text{miss}$ distribution in control regions defined by requiring both AK8 jets to be outside the SR, renormalized to $B_{norm}$. 

Here we point out that we can use one of the $Z$'s to define a SR for resonant anomaly detection, and then we can use the rest of the kinematic variables ($m_{jet}$ of the subleading AK8 jet, $p_T^\text{miss}$, $H_T$, etc) to play the role of $x$ in resonant anomaly detection. We show that this allows for a potentially more expansive and model-agnostic search, while not sacrificing much in sensitivity to the original SUSY signal. We illustrate this with additional SUSY-motivated scenarios (different decay branching ratios to $h$ and $Z$), as well as hypothetical nonminimal scenarios involving non-SM resonances. 

Notably, all of these scenarios have $p_T^\text{miss}$, and in fact the $p_T^\text{miss}$ is essential to suppress the resonant backgrounds from SM $Z/W$+jets with hadronically decaying $Z/W$. This leads to a novel combination of a resonant and nonresonant anomaly detection strategy.\footnote{See \cite{Finke:2022lsu} for a different, fully nonresonant application of weakly-supervised anomaly detection to the jet constituents of the monojet+$p_T^\text{miss}$ final state. Motivated by (nonresonant) dark showers, they did not obtain their background templates from sidebands in the jet mass; instead they considered an idealized (perfect) background template from simulated $Z(\nu\nu)$+jets events.}
This is also the first application of model-agnostic strategies to the SUSY domain and opens up the potential for many more new avenues in the search for SUSY and other well-motivated top-down scenarios. Our method should be contrasted with existing ML-based approaches to SUSY in the literature which are fully supervised (see e.g.\ 
\cite{Mullin:2019mmh,CMS-PAS-SUS-21-006,ATLAS:2022ihe,CMS:2023ktc,Choudhury:2023eje}).

The outline of our paper is as follows: Section \ref{sec:data} describes how the signal and background processes are simulated. In section \ref{sec:CATHODE} we summarize the steps involved in CATHODE. We show the results of applying CATHODE to different signal processes in section \ref{sec:results}. We conclude in section \ref{sec:conclusions}.
Finally, in two appendices we describe our recasting of the LHC analysis, and the CATHODE receiver operating characteristic (ROC) curves for various signal models.
\section{Data}
\label{sec:data}

Since all the methods described here (both the CMS search and CATHODE) fully rely on data for estimating backgrounds (aka are ``fully data-driven"), the simulation data we generate here is meant to play the role of real data, and all background estimates and significances etc we derive are meant to illustrate the result one would get applying these methods to collider data. There will be no events generated here that play the role of simulations at the LHC.

For Standard Model (SM) background data, we take into account the three largest contributions of background events to the CMS search,
arising from $Z$ + jets, $W$ + jets and $t \bar{t}$+ jets. $W$ and $Z$ events were generated with one to four additional final state partons while $t \bar{t}$ were generated with up to 3 additional partons.

For the benchmark signal (to be used to compare the performance of the CMS search vs.\ the CATHODE method), we follow the CMS search and generate gluino pair production (with 0 to 2 additional partons), with subsequent cascade decay $p p \rightarrow \widetilde{g} \widetilde{g}, \widetilde{g}\rightarrow q \bar{q} \widetilde{\chi}_2^0, \widetilde{\chi}_2^0 \rightarrow Z \widetilde{\chi}_1^0$ where the neutralino $\widetilde{\chi}_2^0$ is the next to lightest supersymmetric particle (NLSP) and $\widetilde{\chi}_1^0$ is the lightest supersymmetric particle (LSP). The mass splitting between the gluinos and NLSP is set to 50~GeV while the LSP mass is 1~GeV. This results in soft jets from the first step of the decay and a highly boosted $Z$ boson. The LSP escapes the detector and contributes large amounts of missing energy.

Later we will also consider decays of $\widetilde{\chi}_2^0$ to $X\widetilde{\chi}_1^0$ where 
 
the $X$ is either a Standard Model Higgs boson or a new Higgs boson with mass besides 125~{\rm GeV} like the new Higgs bosons in supersymmetric extensions of the Standard Model. The Standard Model Higgs boson decays in $\sim 58\%$ of cases to $b \bar{b}$ while for the latter case we set the branching ratio to $100\%$.

All events are generated with  {\sc MadGraph5\textunderscore aMC@NLO 3.2.0} with $\sqrt{s}=13~{\rm TeV}$. The {\sc NNPDF3.1LO} PDF set~\cite{NNPDF:2017mvq} is used throughout. At the generator level a minimum $H_T$ cut of 250~GeV is imposed. Gluinos are decayed spin uncorrelated with {\sc Madspin}~\cite{Artoisenet:2012st} to $q \bar{q} \widetilde{\chi}_2^0 $ via an off-shell squark and subsequently $\widetilde{\chi}_2^0 \rightarrow X \widetilde{\chi}_1^0$.
Showering is done using {\sc Pythia 8.306}~\cite{Bierlich:2022pfr} with MLM \cite{MANGANO2002343,MLM2} merging. {\sc Pythia}-Tune CP5 was used for background events while CP2~\cite{CMS:2019csb} was used for the signal samples. The number of background events in each channel is scaled to match their respective next-to-leading-order cross sections~\cite{Alwall:2014hca}.
Detector effects are simulated using {\sc Delphes 3.5.0}~\cite{deFavereau:2013fsa} with the \texttt{delphes\textunderscore card\textunderscore CMS.tcl} detector card modified to account for the lepton isolation criterion. Particles are clustered into jets using the anti-$k_T$ clustering algorithm with cone-radius parameter $R=0.4$ for AK4 jets and $R=0.8$ for AK8 jets. To be considered jets have to have $p_T>30$ and $\vert \eta \vert < 2.4$.

The following selection criteria are imposed for both the classical CMS-recast and the dataset for CATHODE:
\begin{enumerate}
\item $N_{\text{AK4 jet}}\geq 2$
\item $p_T^\text{miss} > 300~{\rm GeV}$
\item $H_T > 400~{\rm GeV}$, where $H_T = \sum_\text{AK4 jets} \vert \vec{p}_T \vert$
\item $\vert \Delta \phi_j , \vec{H}_T^\text{miss}\vert > 0.5 (0.3)$ for the first two (up to next two) AK4 jets, where $\vec{H}_T^\text{miss}=-\sum_\text{AK4 jets} \vec{p}_T$
\item no isolated photon, electron or muon candidate with $p_T > 10~{\rm GeV}$ with isolation variables $I < 0.1 , 0.2 $ and $1.3~{\rm GeV}/p_T +0.005$ for isolated electron, muon and photon respectively
\item no isolated track with\\ $m_T = \sqrt{2 p_T^\text{track} p_T^\text{miss} (1-\cos (\phi^\text{miss}-\phi^\text{track})}<100~{\rm GeV}$ and $p_T > 5~{\rm GeV}$ for tracks identified as an electron/muon or else 10~GeV.
\item at least 2 AK8 jets with $p_T > 200~{\rm GeV}$
\end{enumerate}
The number of background events that pass this baseline selection is shown in the first line of table \ref{tab:eventcounts}. In total, the dataset is composed of 107,421 background events corresponding to $\mathcal{L}_\text{int}=300~{\rm fb}^{-1}$ after cuts 1-7. Signal events are injected according to the gluino-pair production cross section.

\begin{table}[t]
\begin{center}
\begin{tabular}{| c | c c c |}
\hline
Selection & $W$ & $Z$ &$t\bar{t}$ \\
\hline
Baseline selection&73790&25725&7906\\
$m_{J_1} \in [70~{\rm GeV},100~{\rm GeV}]$&5936&2401&1320\\
CMS-SUS-19-013~\cite{CMS:2020fia} signal region&420&237&153\\
\hline
\end{tabular}
\caption{Number of events passing each selection requirement for $\mathcal{L}_\text{int}=300~{\rm fb}^{-1}$ }
\label{tab:eventcounts}
\end{center}
\end{table}

\begin{figure}[t]
\centering
\includegraphics[width=\linewidth]{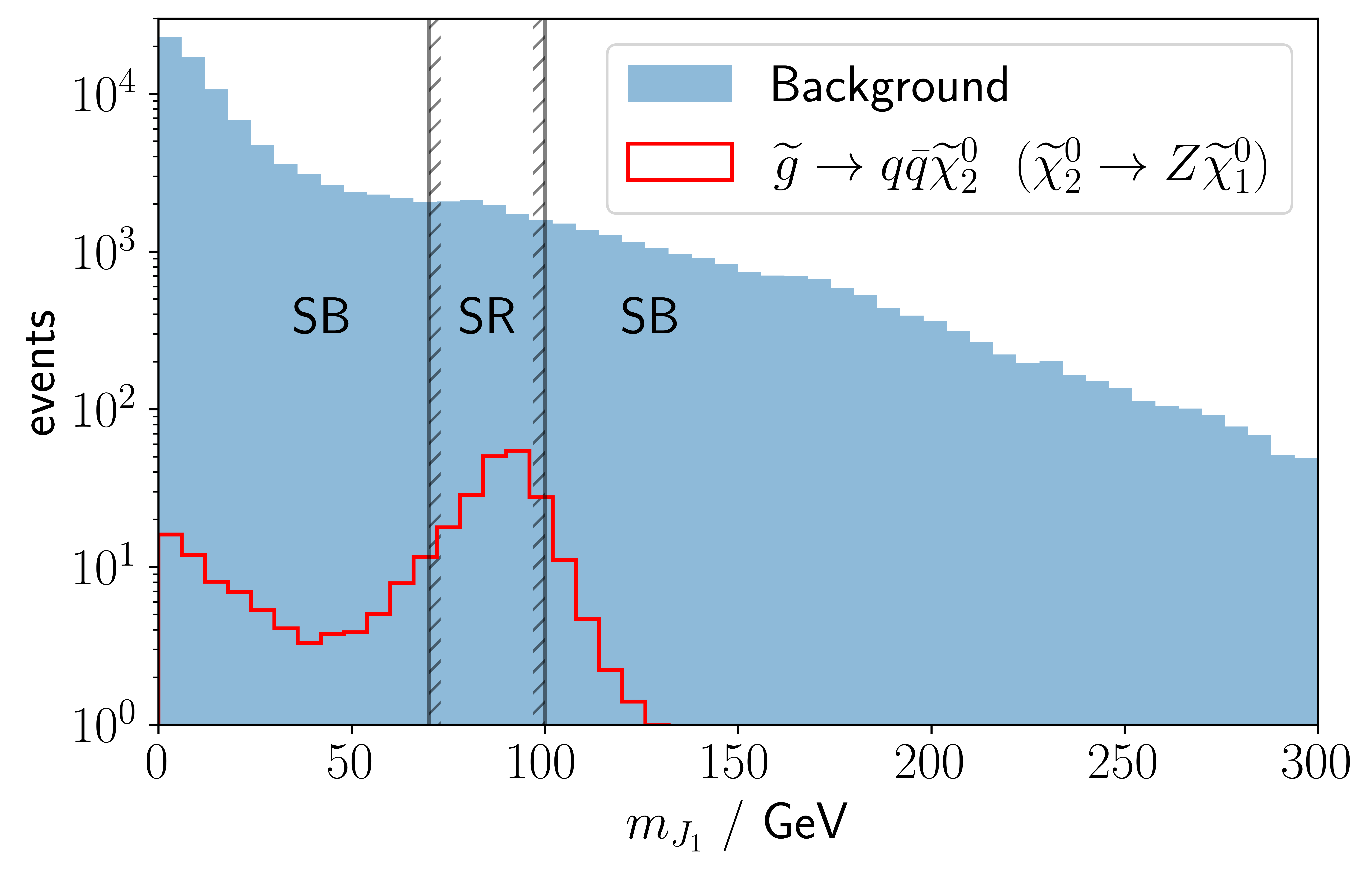}
\caption{Distribution of the resonant feature $m_{J_1}$ for background and signal events in the sideband (SB) and the signal region (SR). The signal corresponds to $m_{\tilde{g}}=1700~{\rm GeV}$. The distributions are scaled to $\mathcal{L}_\text{int}=300~{\rm fb}^{-1}$. }
\label{fig:mj_distribution}
\end{figure}

\begin{figure*}[ht!]
    \centering
    \includegraphics[width=0.49\textwidth]{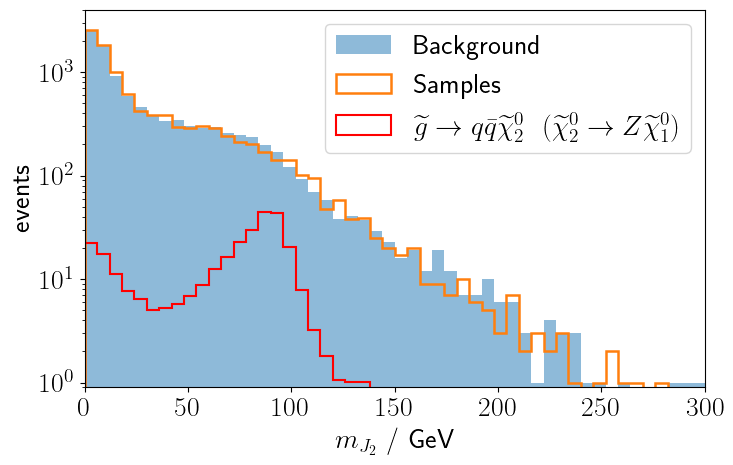}
    \includegraphics[width=0.49\textwidth]{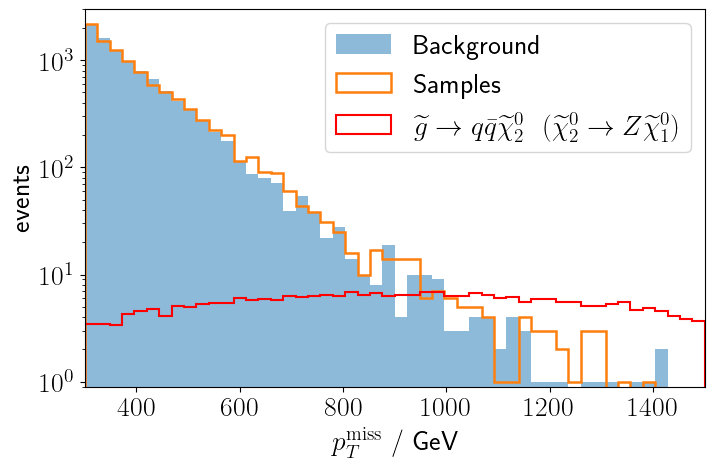}
    \includegraphics[width=0.49\textwidth]{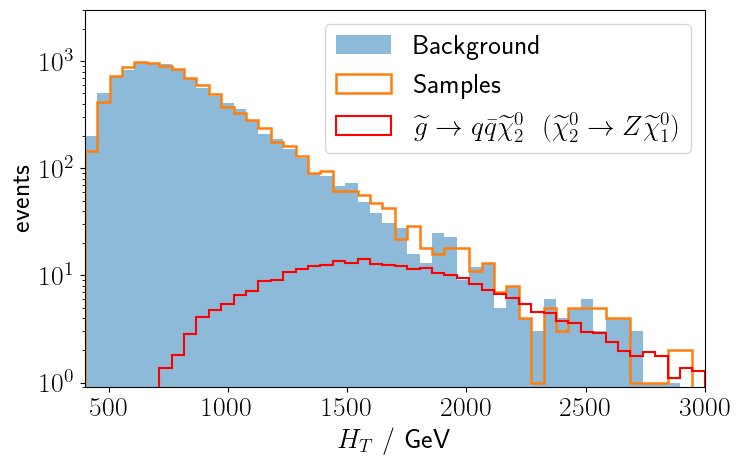}
    \includegraphics[width=0.49\textwidth]{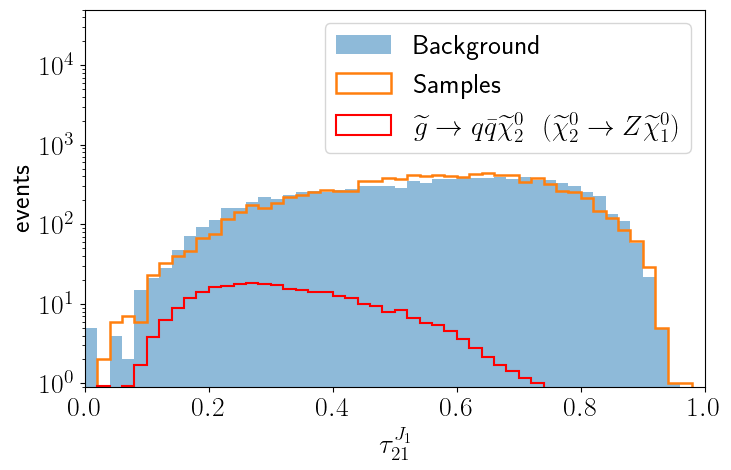}
    \includegraphics[width=0.49\textwidth]{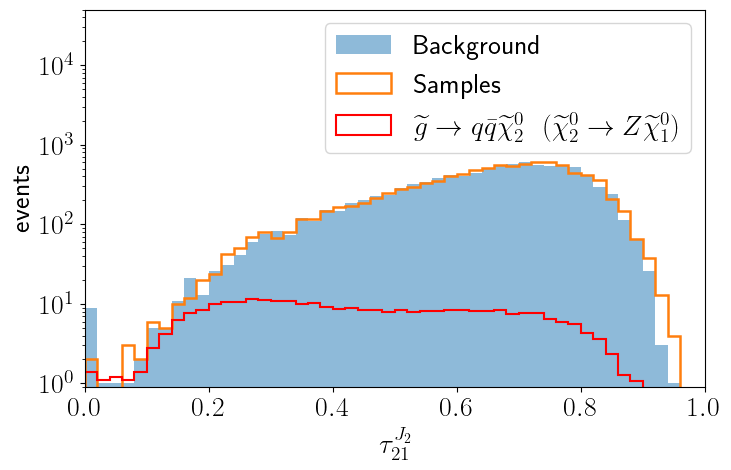}
    \caption{Comparison of the signal and background distribution inside the signal region and the artificial samples. The artificial samples will be discussed in the next section. The signal corresponds to $m_{\tilde{g}}=1700~{\rm GeV}$. The distributions are scaled to $\mathcal{L}_\text{int}=300~{\rm fb}^{-1}$.}
\label{fig:SampledEvents}
\end{figure*}

Figure \ref{fig:mj_distribution} shows that the feature $m_{J_1}$ is smooth for the background while it is resonant for the signal. (Hadronically decaying $W$'s and $Z's$ are eliminated by the requirements on $p_T^\text{miss}$.) This is a necessary feature for the application of the CATHODE method employed in section \ref{sec:CATHODE}. 
Figure \ref{fig:SampledEvents} shows that the signal of new physics is found on the tail of the $p_T^\text{miss}$ distribution, while the background peaks at lower $p_T^\text{miss}$. We will show that the powerful discriminator $p_T^\text{miss}$ can be leveraged by CATHODE even though the signal is found on the tail of the distribution.

\section{CATHODE}
\label{sec:CATHODE}

Here we recap the main points of the inner workings of Classifying Anomalies THrough Outer Density Estimation CATHODE (for more detail see~\cite{Hallin:2021wme}). In very broad strokes, CATHODE aims to learn the density of background events in a signal-depleted region and estimates the density inside the signal enriched region by interpolation. Then, artificial samples are generated in that region, which should follow a signal-depleted distribution.  Using a classifier, which is trained to distinguish between the artificial and real events, we can approximate the likelihood ratio~\eqref{eq:anomalyscore}.
This would be the ideal (optimal) model-agnostic anomaly detector, as it is monotonic with $p_\text{signal}(x) / p_\text{bg}(x)$ for any signal (since $p_\text{data}(x)$ is an admixture of $p_\text{signal}(x)$ and $ p_\text{bg}(x)$)~\cite{Metodiev:2017vrx}.
This allows CATHODE to classify data events as background-like or signal-like. The whole method works by learning directly form data. The training and model selection of both the density estimation and classification is completely agnostic of any signal truth label.

In this study, the events are represented as the tuple $m_{J_1}$ and $x$ with 
\begin{equation}\label{eq:fivedfeatures}
x=\left(m_{J_2}, p_T^\text{miss},H_T, \tau_{21}^{J_1},\tau_{21}^{J_2}\right)
\end{equation} where $J_1$, $J_2$ are the leading/subleading AK8 jets and $\tau_{21}=\tau_2 / \tau_1$ is the ratio of n-subjettiness variables~\cite{Thaler:2010tr}.
To compare the technique to the classical search more directly we also consider the reduced set of features 
\begin{equation}\label{eq:threedfeatures}
x=\left(m_{J_2}, p_T^\text{miss},H_T\right)
\end{equation}
so that CATHODE only gets to use the same information.
We use a slightly modified version of the original repository\footnote{\url{https://github.com/HEPML-AnomalyDetection/CATHODE}} to allow for any dimension for $x$. 

\subsection{Data preparation and density estimation}

First, one defines the signal region (SR) as an interval in $m_{J_1}$ where the signal is expected to be concentrated similar to a classical bump hunt. The complement of the SR defines the sideband (SB). 
As in any bump hunt, the SR window has to account for the position and the width of the signal bump. Because the reconstructed jet mass is not distributed symmetrically around the mass $m$ of the mother particle  (which is the $Z$, the Higgs or a BSM Higgs in this paper), we chose parameterization
\begin{equation} \label{eq:SR}
 m_{J_1} \in [m  \left( 1-\frac{4}{3}\sigma_m\right),m  \left( 1+\frac{2}{3}\sigma_m\right)].
\end{equation}
  We estimate the mass resolution to $\sigma_\text{m} = 15\%$ and round the window to the closest GeV. The lower sideband extends to $m_{J_1}=0$ while the upper sideband is only bound by the phase space.

Events in the SB are partitioned into a training set (75\%) used for the actual training and in a validation set (25\%)¸ used to select the models used in the next steps. To address the finite number of real SB events we use leave-one-out cross validation such that we get four datasets with nonoverlapping validation sets. The data is transformed (preprocessed) for easier learning by shifting and scaling the observables in $x$ to fit the interval $(0,1)$, then applying a logit tranformation\footnote{$\text{logit}(x)=\ln \frac{x}{1-x}$}, and again shifting and scaling to unit standard deviation and zero mean.

For density estimation, a Masked Autoregressive Flow (MAF) is used with affine transformations~\cite{MAF}.
The MAF constructs invertible transformations with tractable Jacobians that map a simple multidimensional distribution (e.g.\ multiple Gaussians as is considered here) to the target density, in this case the conditional probability $p_\text{data}(x\vert m_{J_1}\in{\rm SB})$. The MAF uses 15 blocks of Masked Autoencoder for Distribution Estimation (MADE)~\cite{MADE} to learn the transformations. The number of events it is trained on depends on the signal region but is typically of the order of $10^5$.\footnote{We emphasize that the number of events we are using for training was carefully tuned to match the actual number of events in data expected in ${\mathcal L}=300/{\rm fb}$.} Training is done with the hyperparameters listed in tab.~\ref{tab:DE}.

\begin{table}[t]
\begin{center}
\begin{tabular}{| c c |}
\hline
Hyperparameter & Value \\
\hline
\verb|optimizer| & Adam\\
\verb|epochs| & 100\\
\verb|learning_rate| & $10^{-4}$\\
\verb|batch_norm| & \verb|true|\\
\verb|batch_norm_momentum| & 1\\
\verb|batch_size| & 256\\
\hline
\end{tabular}
\caption{Parameters of the density estimator}
\label{tab:DE}
\end{center}
\end{table}
After training the ten epochs with the lowest validation loss are selected for the sampling step.

\subsection{Sampling SR events}

The next step aims to sample synthetic events inside the SR using the four density estimates of the last step. Kernel density estimation with Gaussian kernel and bandwidth of 0.01 is used to model the $m_{J_1}$ distribution inside the SR. This is then used to sample $N=1,000$ events from each of the ten DE models which are combined, shuffled and split between the training set (60\%) and validation set (40\%) for the next step. The training and validation sets of all four 
density estimators are combined respectively to form the synthetic dataset with a total of 40,000 events. Compared to the roughly 10,000 real events in the SR (see Table~\ref{tab:eventcounts} second line) this is intentionally oversampled to improve the classification performance~\cite{Hallin:2021wme}. Setting $N$ even higher did not improve results systematically.  
The synthetic background events and the real SR events are then standardized in the SR without the logit transformation.

The distributions of the synthetic events are shown in orange in figure \ref{fig:SampledEvents}. In all of our models the signal is located in a resonance in $m_{J_2}$ and in the tail of the $p_T^\text{miss}$ distribution. The density estimation has to model the shape reasonably well so this powerful classification feature can be leveraged. This is accomplished successfully as shown in figure \ref{fig:SampledEvents}.

\subsection{Classifier and anomaly detection}

Now a classifier is trained on both the synthetic and real SR dataset to distinguish the sampled events, which should follow the background distribution, from the real events, which additionally might contain events following the signal distribution.

The classifier consists of 3 hidden layers with 64 nodes and ReLU activation each and it is optimized using the hyperparameters given in tab.~\ref{tab:Classifier}. Because the datasets are imbalanced, a weight is assigned such that both classes contribute equally to the loss.

Since in a realistic example the number of events to train and validate on is limited, we employ an additional step of leave-one-out cross validation. The real SR data is partitioned into four subsets of equal size. In each subset, one quarter of the real events are held back as a test set for the anomaly detection while the remaining 75\% are split between the training set (60\%) and the validation set (40\%). (The synthetic background events are also split into train/val sets with the same proportions.) After training, the ten model states with the lowest validation loss are selected and evaluated on the test set. The predicted labels are then averaged over the models and assigned as anomaly scores to the events. This is repeated for the next quarter of the SR data, and so on, until every event in the SR is assigned an anomaly score.

\begin{table}[t]
\begin{center}
\begin{tabular}{| c c |}
\hline
Hyperparameter & Value \\
\hline
\verb|optimizer| & Adam\\
\verb|epochs| & 100\\
\verb|learning_rate| & $10^{-3}$\\
\verb|batch_size| & 128\\
\hline
\end{tabular}
\caption{Parameters of the classifier}
\label{tab:Classifier}
\end{center}
\end{table}

\begin{figure}[b]
\centering
\includegraphics[width=0.9\linewidth]{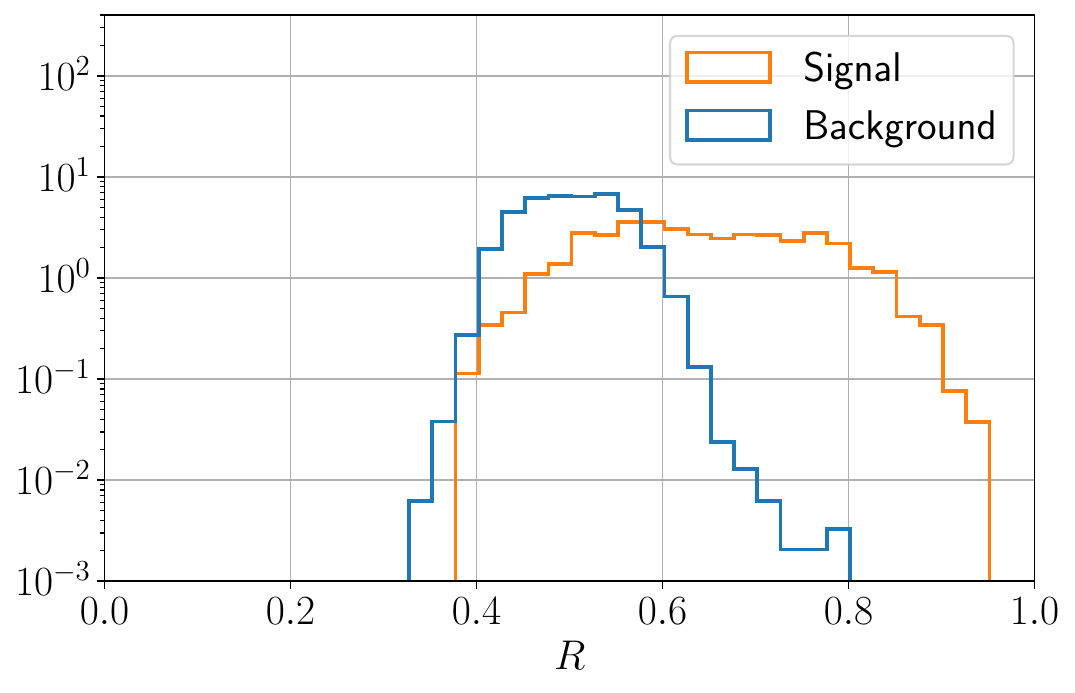}
\caption{Normalized distributions of the anomaly score $R$ of the signal and background processes. The signal corresponds the average distribution of ten independent injections with $m_{\tilde{g}}=1900~{\rm GeV}$.}
\label{fig:R}
\end{figure}

To reduce the statistical effects of severely overperforming and underperforming models, each dataset is shuffled 5 times to allow different selections. Then the entire process of the preceding paragraph is repeated to produce 5 different anomaly scores. All 5 anomaly score assignments are averaged to produce a final, more robust score. 

Finally, to even out the influence of signal-event selection, everything is repeated ten times with differing independent sets of signal events. In all the results we report below, we will report the mean and standard deviation of these ten different trials.

The signal to background ratio is improved by cutting on the anomaly score above a critical value $R_c$. Figure \ref{fig:R} shows the distributions of the anomaly score $R$ for the signal and background. No additional selections are performed. In a real application one would perform statistical inference by means of a bump hunt on the $R$ distribution which is beyond the scope of this work. Instead the performance is evaluated using the nominal significance $\text{Z}=S/\sqrt{B}$ with $S$ ($B$) the number of signal (background) events after imposing this cut. This makes use of the truth labels which an experiment would have to replace by other means of background estimation.
One still has to chose a strategy to set $R_c$. In the following we will show the signal significance with $R_c$ set to maximize Z with at least 5 background events left to show the best performance one could hope for. Since a real application does not have access to the truth labels this is not immediately applicable. To show a more realistic method we also show the performance where $R_c$ is set so that 1\% of SR events pass the cut while also containing at least 5 background events.

\section{Results}
\label{sec:results}

\subsection{Nominal signal model}

We first turn our attention to the nominal signal model where $\widetilde{\chi}_2^0 \rightarrow Z \widetilde{\chi}_1^0$. This is the signal model the dedicated CMS search~\cite{CMS:2020fia} was aimed at.

\subsubsection{Three features}

We start by using the limited feature set $x=\left(m_{J_2}, p_T^\text{miss},H_T\right)$ so CATHODE does not have access to more information than the classical search.
To compare with CMS we calculate the signal significance for events inside the signal region $m_{J_1 / J_2} \in [70~{\rm GeV},100~{\rm GeV}]$ with the b-veto mentioned in section \ref{sec:recreatingCMS} applied. Since the search gets most of its sensitivity from the highest $p_T^\text{miss}$-bins, we apply an additional cut $p_T^\text{miss}> 800~{\rm GeV}$.\footnote{Technically, the original CMS search uses $p^\text{miss}_T$-bins, and most of the sensitivity comes from the three highest bins, 800--1000~GeV, 1000--1200~GeV and larger than 1200~GeV, where the background is comparable or subdominant to the signal hypothesis. 
To get a fair comparison with CATHODE we replace this with a single cut.} This leads to roughly the same number of events as when only the top 1\% of events are kept for CATHODE.
For a gluino mass with sizable cross section like 1700~GeV the classical search yields on average for ten independent signal injections $\text{Z}=20$. Using CATHODE with 3 features the significance is on average $\text{Z}=34\pm 2$. 

Evidently, CATHODE outperforms the classical approach, even though CATHODE is more model agnostic. The reason is that the classical approach, being cut based,  misses correlations between the features that the multivariate classifier of CATHODE can pick up.

To confirm this, we also investigated the sensitivity of a fully supervised approach, using the same classifier architecture and hyperparameters as that of CATHODE. The training data for the fully supervised classifier consists of an additional 300~fb$^{-1}$ background events and 10,000 signal events. 60\% of this dataset is used in training while the remaining 40\% is used as a validation set to select the best performing model. Evaluating this classifier again with selecting only the top 1\% of anomaly scores results in a significance of on average $\text{Z}=33\pm 4$.  We conclude that CATHODE is saturating the performance of the fully-supervised classifier for this amount of signal (unsurprisingly, since this is a lot of signal), and that
the deep neural networks of both CATHODE's classifier and the supervised classifier can  leverage correlations to improve the signal significance significantly over the classical approach.

\subsubsection{Five features}

\begin{figure}[t!]
\centering
\includegraphics[width=0.9\linewidth]{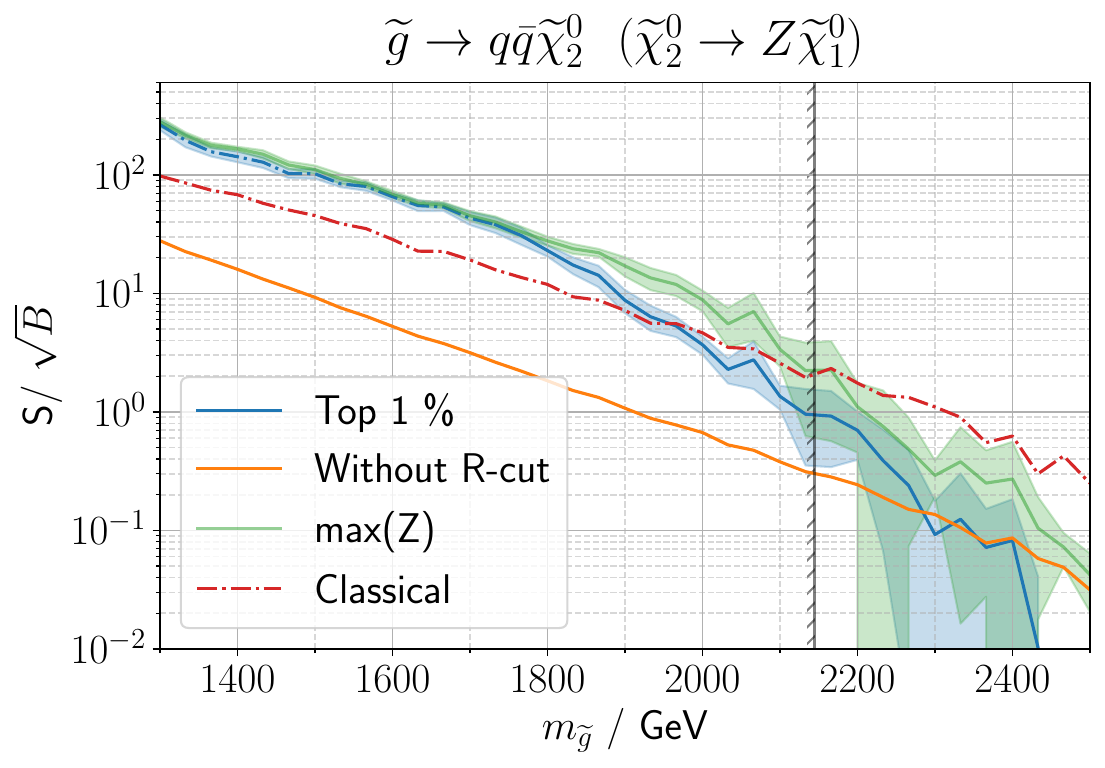}
\caption{Sensitivity of CATHODE and the classical strategy. The signal window is set as $m_{J_1} \in [70~{\rm GeV},100~{\rm GeV}]$. For the blue line $R_c$ is set to allow 1\% of events to pass this cut while the orange line omits the cut completely. The shaded region shows one standard deviation around the mean $S/\sqrt{B}$ obtained from ten different signal injections. The dot-dashed part of the blue line represents parameter points where $R_c$ has to be lowered to allow 5 background events. The vertical black line at 2145~GeV indicates gluino mass that is excluded at 95\% confidence level by our 300/fb recreation of the dedicated search~\cite{CMS:2020fia} . The red dot-dashed line is calculated using the classical  strategy with $m_{J_1 / J_2} \in [70~{\rm GeV},100~{\rm GeV}]$, $p_T^\text{miss} > 800~{\rm GeV}$ and the b-veto.}
\label{fig:ZZ}
\end{figure}

From now on we will use the five features $\left(m_{J_2}, p_T^\text{miss},H_T, \tau_{21}^{J_1},\tau_{21}^{J_2}\right)$ because the subjettiness variables $\tau_{21}$ are useful discriminants. Figure~\ref{fig:ZZ} shows CATHODE's performance compared to the classical strategy. We see that in the relevant region at high gluino masses the conservative cut on $R$ (allowing only the top 1\% to pass) reaches only slightly weaker results. We identify the mass where the signal significance is $\text{Z}=1.645$ with the expected 95\% limit on the mass in a real application \cite{Bhattiprolu_2021}. The conservative cut on $R$ alone excludes gluino masses up to $m_{\widetilde{g}}=2066~{\rm GeV}$. This is only slightly weaker than the expected excluded mass of $m_{\widetilde{g}}<2145~{\rm GeV}$ for a dedicated search at this integrated luminosity. This is expected because a model specific search will be fine-tuned to the specific process while CATHODE is intentionally kept more general. CATHODE's strength lies in this generalization as it is able to detect different models without the need to tweak the approach as we will show in the following sections.

\subsection{Alternate signal model: decays to SM Higgs}

\begin{figure}[b]
\centering
\includegraphics[width=0.9\linewidth]{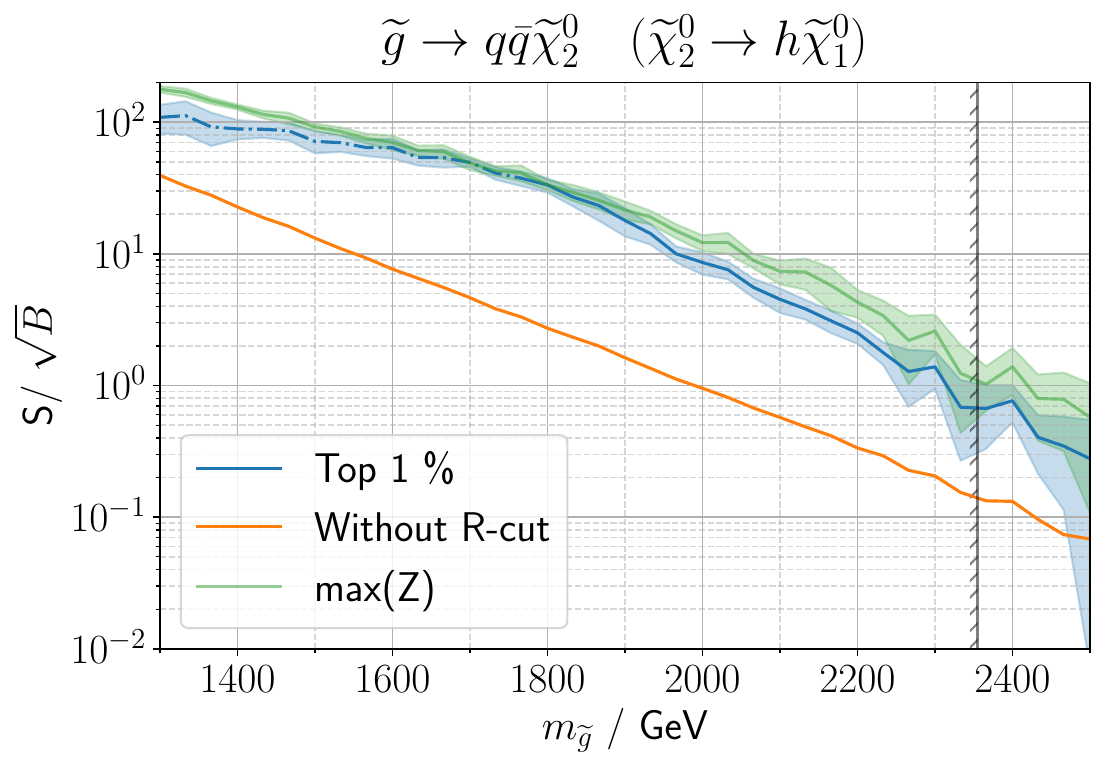}
\caption{CATHODE's performance for $\widetilde{\chi}_2^0 \rightarrow h \widetilde{\chi}_1^0$. The signal window is set as $m_{J_1} \in [100~{\rm GeV},140~{\rm GeV}]$. For the blue line $R_c$ is set to allow 1\% of events to pass this cut while the orange line omits the cut completely. The dot-dashed part of the blue line represents parameter points where $R_c$ has to be lowered to allow 5 background events. The shaded region shows one standard deviation around the mean $S/\sqrt{B}$ obtained from ten different signal injections. The vertical black line at 2355~{\rm GeV} indicates gluino mass that is expected to be  excluded by rescaling the (expected) limit from a dedicated CMS search for this decay~\cite{CMS:2022vpy} from 137/fb to 300/fb integrated luminosity. There is no red line corresponding to the classical search (as in Fig.~\ref{fig:ZZ})  because we did not perform a detailed recast of~\cite{CMS:2022vpy}. }
\label{fig:HH}
\end{figure}

Now we turn our attention to another model, where the neutralinos decay via $\widetilde{\chi}_2^0 \rightarrow h \widetilde{\chi}_1^0$ where $h$ is the 125~GeV Standard Model Higgs boson. All that has to be done for CATHODE is select a new signal window around 125~GeV. A scan over the gluino mass is shown in figure \ref{fig:HH}. A b-jet selection criterion would be beneficial in this case, but we omit this to keep CATHODE as general as possible. Even without the b-tag CATHODE still generates a sizable signal significance for gluino masses comparable to the expected excluded value. While the dedicated search is expected to exclude gluino masses below 2355 GeV, CATHODE with the 1\% cut reaches $\text{Z}>1.645$ for all masses up to 2233 GeV. With the best possible cut on $R$ this can be pushed to 2300 GeV. As expected CATHODE results in slightly weaker bounds. The opportunity cost of this is significantly lower than a specialized search. The only change in the approach is the choice of the signal region. The intended use of CATHODE scans the signal region over the entire mass range, such that both the decay to $Z$ and Higgs bosons would be included automatically in this strategy without any extra considerations.

\subsection{Alternate signal model: mixed \texorpdfstring{$Z/h$}{Z/h} decays}

\begin{figure}[b]
\centering
\includegraphics[width=0.9\linewidth]{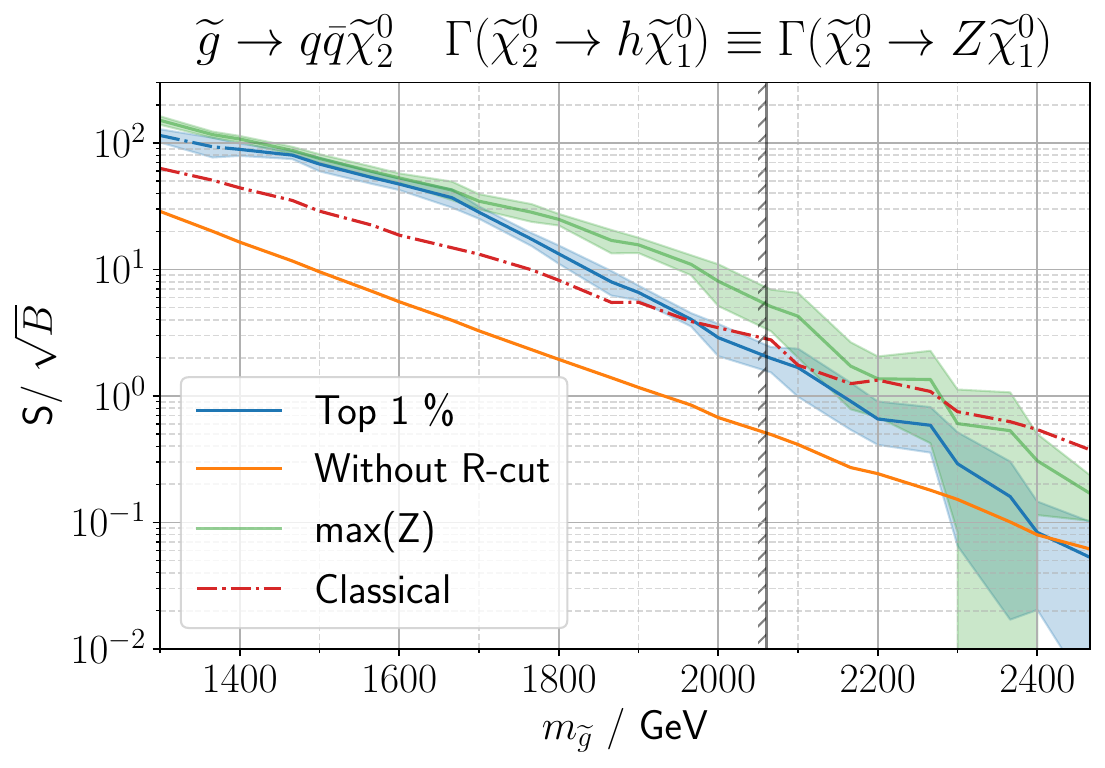}
\caption{Sensitivity of CATHODE and the classical strategy. The signal window is set as $m_{J_1} \in [70~{\rm GeV},140~{\rm GeV}]$. For the blue line $R_c$ is set to allow 1\% of events to pass this cut while the orange line omits the cut completely. The dot-dashed part of the blue line represents parameter points where $R_c$ has to be lowered to allow 5 background events. The shaded region shows one standard deviation around the mean $S/\sqrt{B}$ obtained from ten different signal injections. The vertical black line at 2060~GeV  indicates gluino mass that is expected to be excluded by rescaling the expected excluded cross section obtained by the dedicated CMS search for this decay~\cite{CMS:2017may} from 35.9/fb to 300/fb integrated luminosity.}
\label{fig:HZ}
\end{figure}

Setting the branching ratio of the $\widetilde{\chi}_2^0 \rightarrow h \widetilde{\chi}_1^0$ or $\widetilde{\chi}_2^0 \rightarrow Z \widetilde{\chi}_1^0$ decays to 100\% is a rather unnatural choice. Therefore we also show CATHODE's performance for a model where both branching ratios are 50\%. This time the anomaly detection has to find two bumps simultaneously. For this we chose the signal window to contain both resonances: $m_{J_1} \in [70~{\rm GeV},140~{\rm GeV}]$. The results of a scan over the gluino masses is shown in figure \ref{fig:HZ}. This time CATHODE seems to outperform the extrapolated bound from the dedicated search~\cite{CMS:2017may}. The extrapolation from 35.9/fb to 300/fb integrated luminosity is quite far and should be taken with a grain of salt. The dedicated search classifies events in 0,1 and 2 Higgs categories using b-tags. The signal model populates all categories simultaneously. The approach using CATHODE only uses the single signal region without further thought to generate these results.

\begin{figure}[t]
\centering
\includegraphics[width=0.9\linewidth]{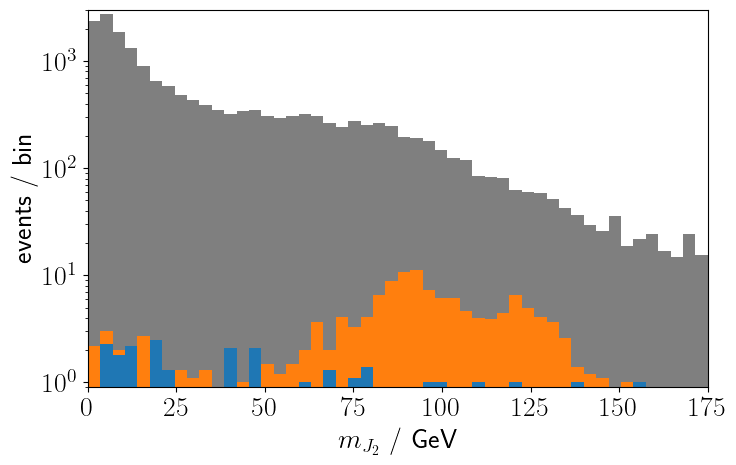}
\caption{The distribution of the data inside the signal region before the anomaly score cut is shown in gray. After selecting the top 1\% of events in the SR the remaining signal events are shown in orange while the remaining background events are shown in blue. The signal corresponds to $m_{\tilde{g}}=1700~{\rm GeV}$. }
\label{fig:ZH-SignalShape}
\end{figure}

In figure \ref{fig:ZH-SignalShape} we show that CATHODE is indeed capable of recovering both bumps corresponding to the decay into $Z$ and Higgs bosons respectively.

\begin{figure}[b]
\centering
\includegraphics[width=0.9\linewidth]{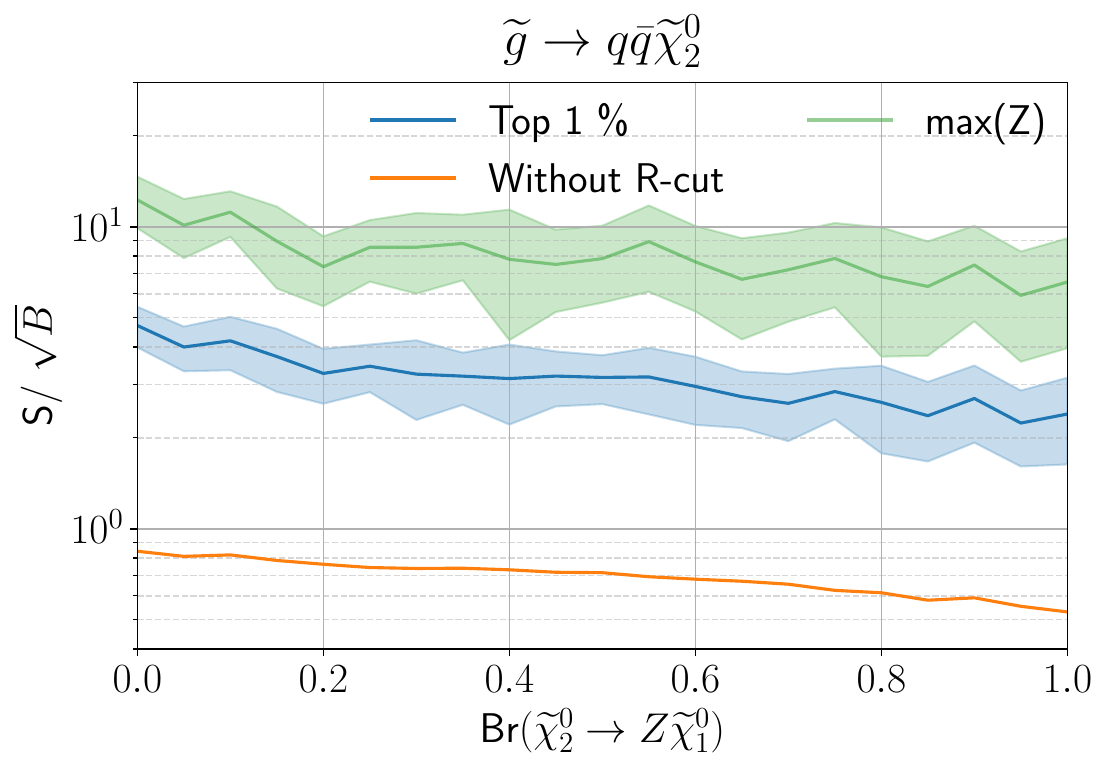}
\caption{Sensitivity of CATHODE for varying branching ratios to $Z$ bosons for $m_{\tilde{g}}=2000~{\rm GeV}$. The shaded region shows one standard deviation around the mean $S/\sqrt{B}$ obtained from ten different signal injections. }
\label{fig:ZH-BR}
\end{figure}

Figure \ref{fig:ZH-BR} shows that CATHODE is very robust against changed in branching ratios. We vary the branching ratio $\text{Br}( \widetilde{\chi}_2^0 \rightarrow Z \widetilde{\chi}_1^0)$ with $\text{Br}( \widetilde{\chi}_2^0 \rightarrow h \widetilde{\chi}_1^0)=1-\text{Br}( \widetilde{\chi}_2^0 \rightarrow Z \widetilde{\chi}_1^0)$ and calculate the significance. Regardless of branching ratio, the multiplicative gain of significance by applying the technique is always between 5 and 6. This shows the real strength of the CATHODE approach over the dedicated searches~\cite{CMS:2020fia,CMS:2022vpy,CMS:2017may}. With the enlarged SR that covers both decay modes, CATHODE only needs to be trained once, independent of the assumption on the BRs, compared to performing a dedicated analysis for each BR assumption. 

\subsection{Alternate signal model: decays to BSM Higgs}

\begin{figure}[t]
\centering
\includegraphics[width=0.9\linewidth]{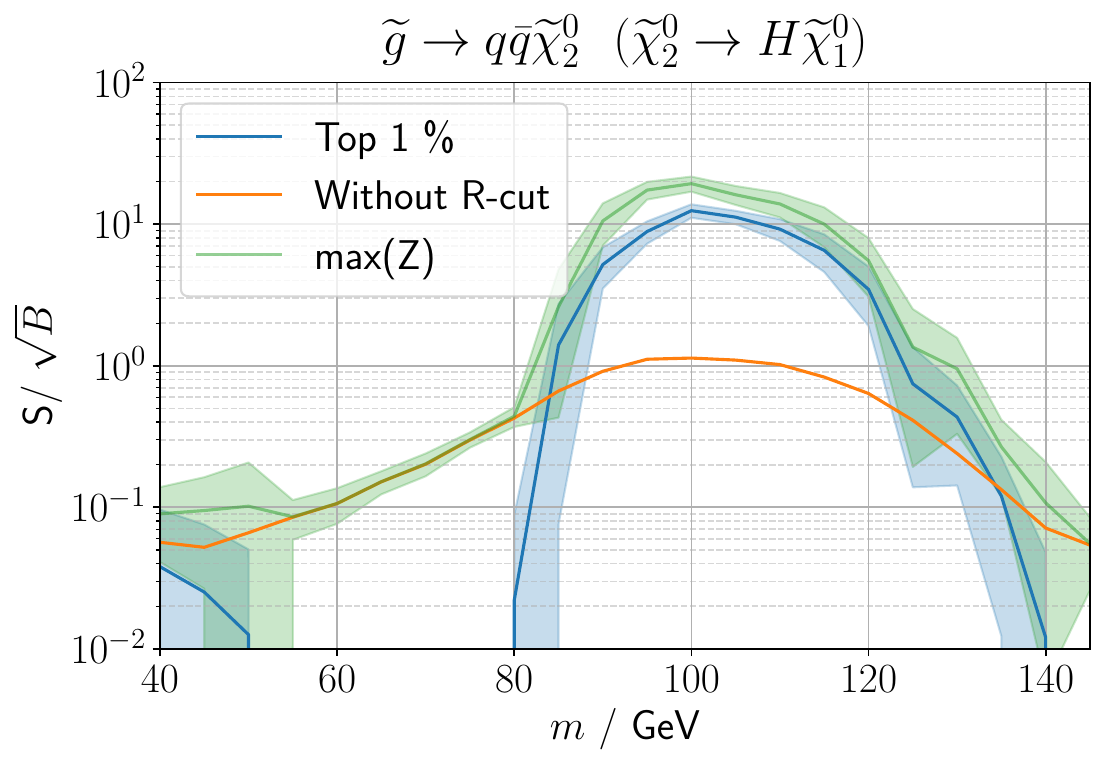}
\caption{Significance for a parameter scan over the mass hypothesis in 5 GeV steps, when the mass is not known a priori. The shaded region shows one standard deviation around the mean $S/\sqrt{B}$ obtained from ten different signal injections. Masses are chosen as $m_{\widetilde{g}}=2000~{\rm GeV}$ and $m_H = 100~{\rm GeV}$.}
\label{fig:unknownMass}
\end{figure}

Until now we applied CATHODE only to models where the position of the bump is known beforehand. But one strength of the technique is that we do not even need to know that. To discuss this further we now focus on another model that induces the neutralino decay $\widetilde{\chi}_2^0 \rightarrow H \widetilde{\chi}_1^0$ where $H$ is one of the additional Higgs bosons introduces by the (N)MSSM that has a mass different from 125~GeV. Because the decay of $H$ depends on the specific implementation of SUSY-breaking parameters we set the branching ratio $BR(H\rightarrow b \bar{b})=100\%$.
To find the signal, CATHODE is applied to different signal regions given by varying mass hypotheses $m$ in equation \ref{eq:SR}, scanning the entire mass range in discrete steps and the signal significance is determined. To demonstrate this we chose $m_H = 100~{\rm GeV}$ and $m_{\widetilde{g}}=2000~{\rm GeV}$ and show the result in figure \ref{fig:unknownMass}. Once the signal window has significant overlap with the signal bump the signal significance gets sufficiently improved to show the presence of anomalous events. In a real application this would then warrant further investigation with a dedicated search.

\begin{figure}[b]
\centering
\includegraphics[width=0.9\linewidth]{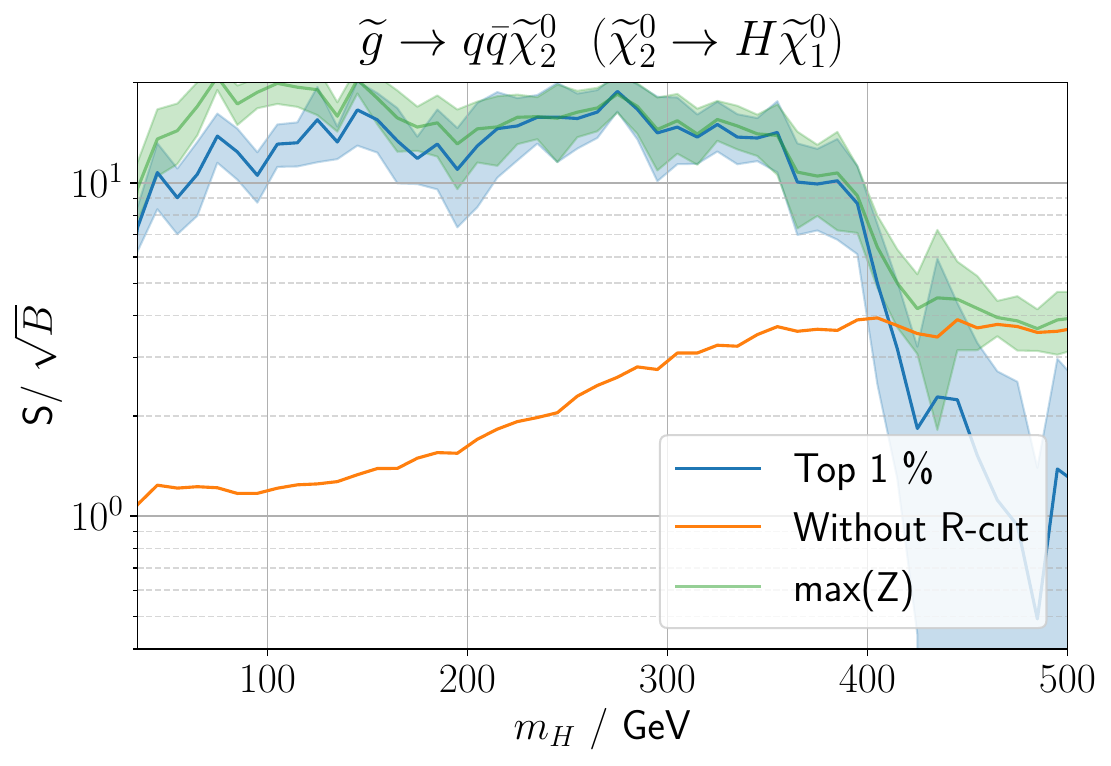}
\caption{Parameter scan of $m_H$ with $m_{\widetilde{g}}=2000~{\rm GeV}$ to show which signals CATHODE can help find in the dataset. The shaded region shows one standard deviation around the mean $S/\sqrt{B}$ obtained from ten different signal injections. }
\label{fig:mhscan}
\end{figure}

Finally we show how wide the possible choice of $m_H$ is that CATHODE can still help to find in our dataset with the given choice of features. For this we perform a parameter scan over $m_H$ from 35~GeV to 515~GeV in 10~GeV steps shown in figure \ref{fig:mhscan}. The method reaches reliably signal significances of order ten up to $m_H \sim 350~{\rm GeV}$ without using b-tags as otherwise powerful discriminators.

\section{Conclusions}
\label{sec:conclusions}

In this paper, we have shown how recently developed techniques for weakly supervised resonant anomaly detection can be easily extended to cover anomalies that also live on tails of distributions. This situation commonly arises in well-motivated weak-scale scenarios such as SUSY, where the cascade decays of heavier BSM particles can produce resonances such as $Z$'s and Higgs bosons, while simultaneously populating the tails of features such as $p_T^\text{miss}$ and $H_T$. As long as the signal is localized in one feature where the background is smooth, resonant anomaly detection can be brought to bear on these additional features in order to enhance the sensitivity to signal.

As a proof-of-concept demonstration, we applied the state-of-the-art anomaly detection method CATHODE~\cite{Hallin:2021wme} to the SUSY scenario $p p \rightarrow \widetilde{g} \widetilde{g}, \widetilde{g}\rightarrow q \bar{q} \widetilde{\chi}_2^0,\widetilde{\chi}_2^0 \rightarrow X \widetilde{\chi}_1^0$ where $X$ is either a $Z$ boson, Standard Model Higgs, or an additional (N)MSSM Higgs boson. Despite being model agnostic, we showed that the CATHODE method is competitive with existing, dedicated, cut-based searches~\cite{CMS:2020fia,CMS:2022vpy,CMS:2017may}, because --- being inherently multivariate --- it takes advantage of correlations between features. Moreover, whereas each decay scenario required a separate, optimized analysis, CATHODE  --- being model agnostic --- is able to simultaneously target them all. 

In this work we considered two different feature sets for the CATHODE algorithm, as shown in eqs.~\eqref{eq:fivedfeatures} and~\eqref{eq:threedfeatures}. These were motivated by the SUSY scenarios we considered, and it would be interesting to generalize our study beyond these feature sets, both to 
increase the degree of model agnosticness of the method, and possibly to enhance the sensitivity to the SUSY signals considered here. For example, our benchmark signals all come with $\sim 4$ additional jets from the gluino decay, and their detailed kinematic distributions (instead of just the aggregate feature $H_T$) may offer additional discriminating power versus the QCD background. Adding features related to  additional jets in the event may also give us more  sensitivity to spectra not explicitly considered here, for example where the NLSP mass is not so close to the gluino mass. As long as $m_{\rm LSP} + m_Z \ll m_{\tilde g}$, the $Z$ will still be boosted, but the extra jets will get harder as $m_{\rm LSP}$ moves away from $m_{\tilde g}$. %Finally, adding more features will make this application of CATHODE more model-agnostic. 

All in all, using modern methods for resonant anomaly detection such as CATHODE allows for a broader and more efficient coverage of the parameter space of physics beyond the Standard Model. With much more data on the way, methods like these should prove indispensable for maximizing the discovery potential of the LHC.

\section{Acknowledgments}

DS is supported by DOE grant DOE-SC0010008. CK would like to thank the Baden-W\"urttemberg-Stiftung for financing through the program \textsl{Internationale Spitzenforschung}, pro\-ject \textsl{Uncertainties – Teaching AI its Limits} (BWST\_IF2020-010).
GK acknowledges support by the Deutsche Forschungsgemeinschaft under Germany’s Excellence Strategy – EXC 2121  Quantum Universe – 390833306.

\appendix

\section{Recasting CMS}
\label{sec:recasting_cms}

We describe how the background samples were simulated as closely as possible to an existing search and verified. We recreate the CMS SUSY search CMS-SUS-19-013~\cite{CMS:2020fia}.
\subsection{Recreating CMS-SUS-19-013}
\label{sec:recreatingCMS}
\begin{figure}[h!]
\centering
\includegraphics[width=0.9\linewidth]{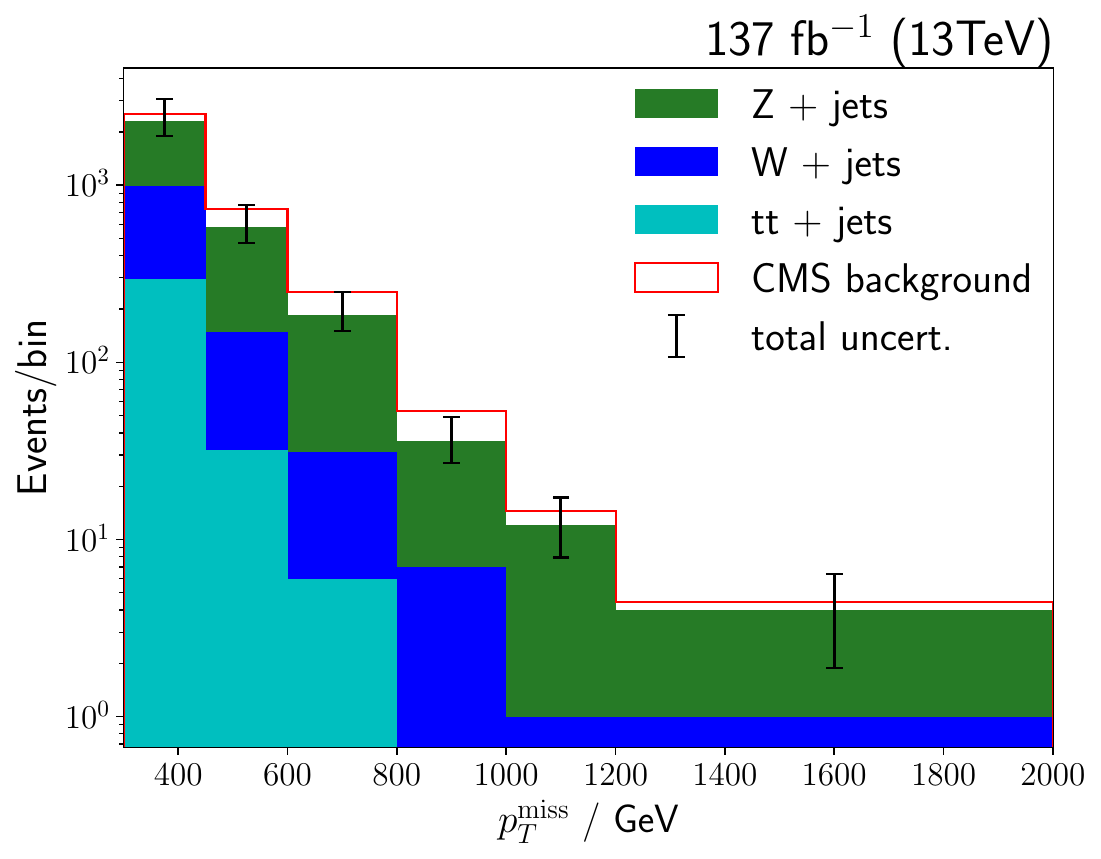}
\caption{$p_T^\text{miss}$ spectrum of the three leading background processes. The background of the same three processes from the CMS publication is shown in red. The variation of cross section due to changing the energy scale by a factor of 1/2 and 2 as computed by MadGraph is assigned as a systematic uncertainty and added to the statistic errors in quadrature and shown as the error bars.}
\label{fig:ptmissCMS}
\end{figure}

The recreation of CMS-SUS-19-013~\cite{CMS:2020fia} follows the most important analysis steps of the original publication. The number of events is set to the integrated luminosity of $\mathcal{L}_\text{int}=137~{\rm fb}^{-1}$. First, a set of remaining cuts are applied to select $Z$ candidates, then the background estimation is recreated before the statistical analysis is performed.
The following cuts are applied to select hadronically decaying $Z$ bosons:
\begin{enumerate}
\setcounter{enumi}{7}
\item Softdropped $m_\text{jet}\in [40~{\rm GeV},140~{\rm GeV}]$ of the two highest $p_T$ AK8 jets
\item $\Delta R_{Z,b}>0.8$ for the second highest $p_T$ AK8 jet $Z$ and any b-tagged jet where the angular separation is defined as $\Delta R=\sqrt{\Delta \phi^2+\Delta \eta^2}$
\end{enumerate}
The resulting $p_T^\text{miss}$ spectrum is shown in figure \ref{fig:ptmissCMS} which agrees with the spectrum shown in the original publication within uncertainties.

The background estimation consists of the normalization and the shape estimation.
The signal region (SR) is defined as $m_\text{jet} \in [70~{\rm GeV},100~{\rm GeV}]$. First one demands the subleading AK8-jet to be in the SR. Then a linear function is fitted to the $m_\text{jet}$ spectrum of the leading AK8 jet outside its SR. The nominal yield $\mathcal{B}_\text{norm}$ is obtained by integrating the linear function in the SR. The statistical error of the yield is obtained from the spread of pseudo-experiments sampled from the fit.
Additionally to the linear function Chebychev functions up to the fourth order are fitted. The largest deviation of the nominal yield is then assigned as an additional uncertainty.

The background $p_T^\text{miss}$ shape is obtained by the sideband (SB) with both AK8 jets outside the SR. The content of the  $i$th $p_T^\text{miss}$ bin is denoted as $N_i^\text{SB}$. The transfer factor from the SB to the SR is then calculated as
\begin{equation}
\mathcal{T}\equiv \frac{\mathcal{B}_\text{norm}}{\sum_i N_i^\text{SB}} = 0.206 \pm 0.023,
\end{equation}
which agrees with the original publication within uncertainties.
The expected background in bin $i$ is then 
\begin{equation}
\mathcal{B}_i = \mathcal{T}N_i^\text{SB}.
\end{equation}

\begin{figure}[h]
\centering
\includegraphics[width=0.9\linewidth]{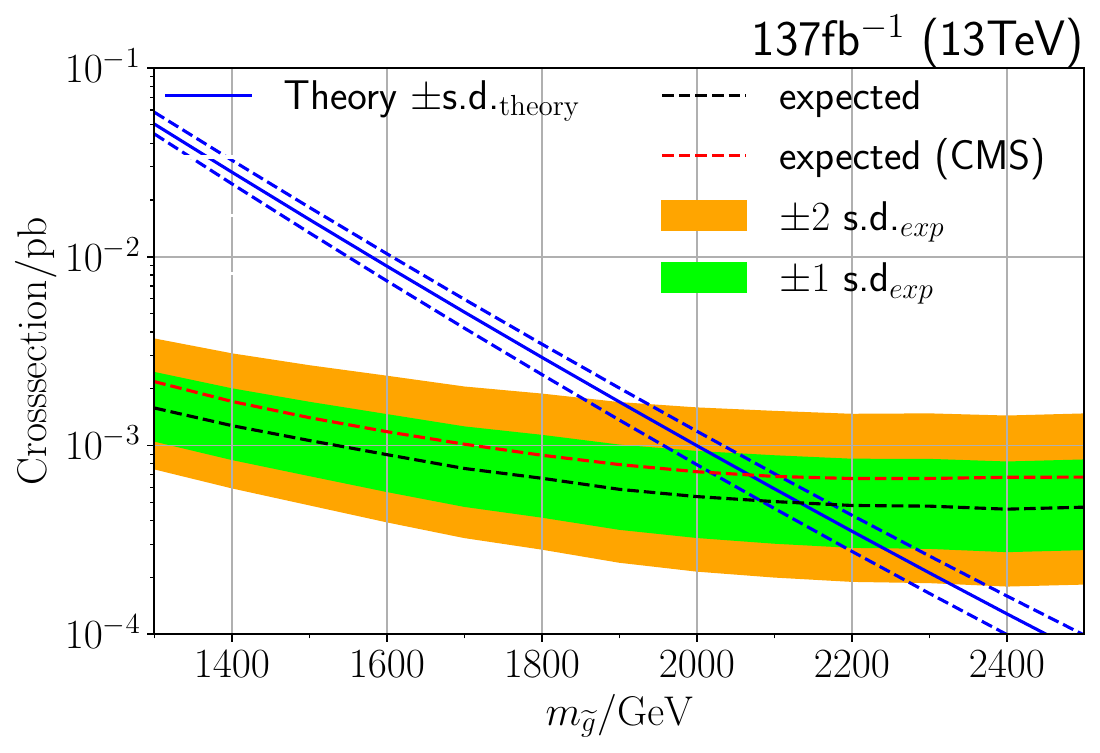}
\caption{Recreation of CMS-SUS-19-013~\cite{CMS:2020fia}. The red dashed line denotes the expected limits of original CMS search. The black dashed line shows the expected limits of the recreation.}
\label{fig:results}
\end{figure}

\begin{figure}[h]
\centering
\includegraphics[width=0.9\linewidth]{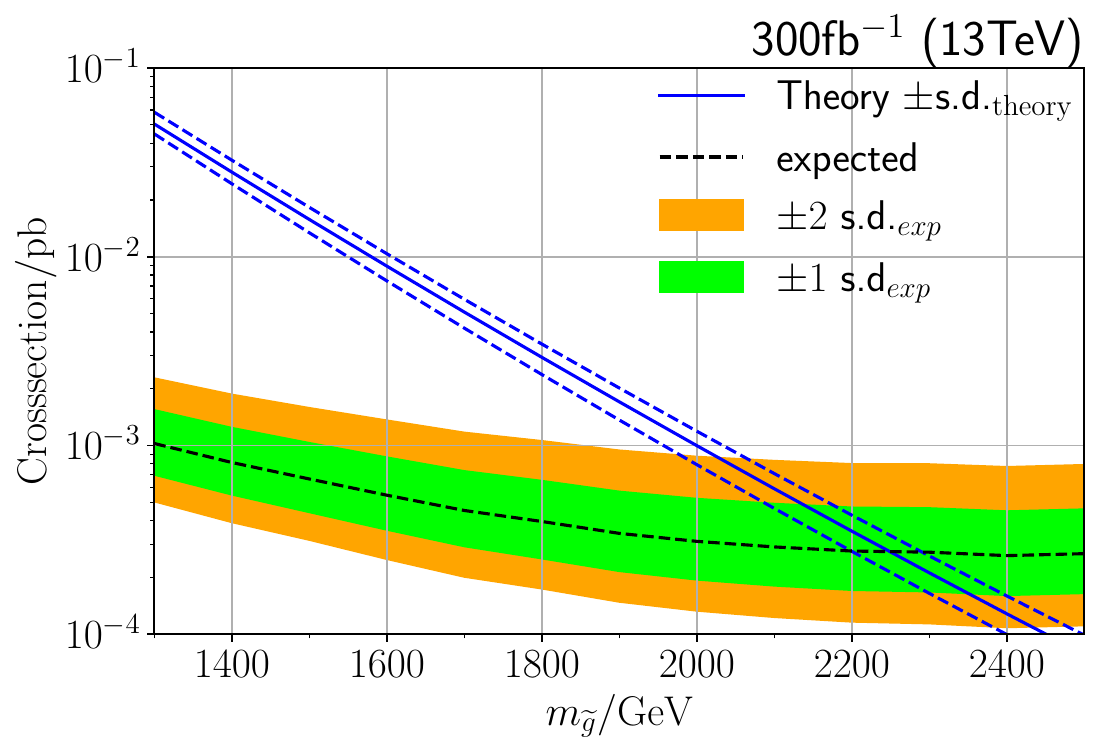}
\caption{Results of the classical search for 300/fb  integrated luminosity}
\label{fig:result300}
\end{figure}

{\sc RooStats}~\cite{rene_brun_2019_3895860} is used for statistical modeling. It takes $N_i^\text{SB}$ with statistical errors, $\mathcal{T}$ and $\Delta \mathcal{T}$  to model the background in the SR with uncertainties. The signal model contains signal events that pass all cuts and is rescaled to the approximate NNLO+NNLL cros section~\cite{Borschensky:2014cia}. The overall uncertainty of the cros section is applied to all signal bins. The resulting statistical model is then evaluated with the CL$_s$ approach and the asymptotic form of the onesided profile likelihood test statistic. This is used to obtain the 95\% C.L. cros sections. The limits are shown in figure \ref{fig:results} for the integrated luminosity $\mathcal{L}_\text{int}=137~{\rm fb}^{-1}$ and in figure \ref{fig:result300} for $\mathcal{L}_\text{int}=300~{\rm fb}^{-1}$. We use the latter dataset for the application of the ML technique since the accuracy is greatly improved with more data points to learn on while in reach for the collider in the near future.

\section{ROC Curves}
\begin{figure*}[ht!]
\centering
\includegraphics[width=0.49\linewidth]{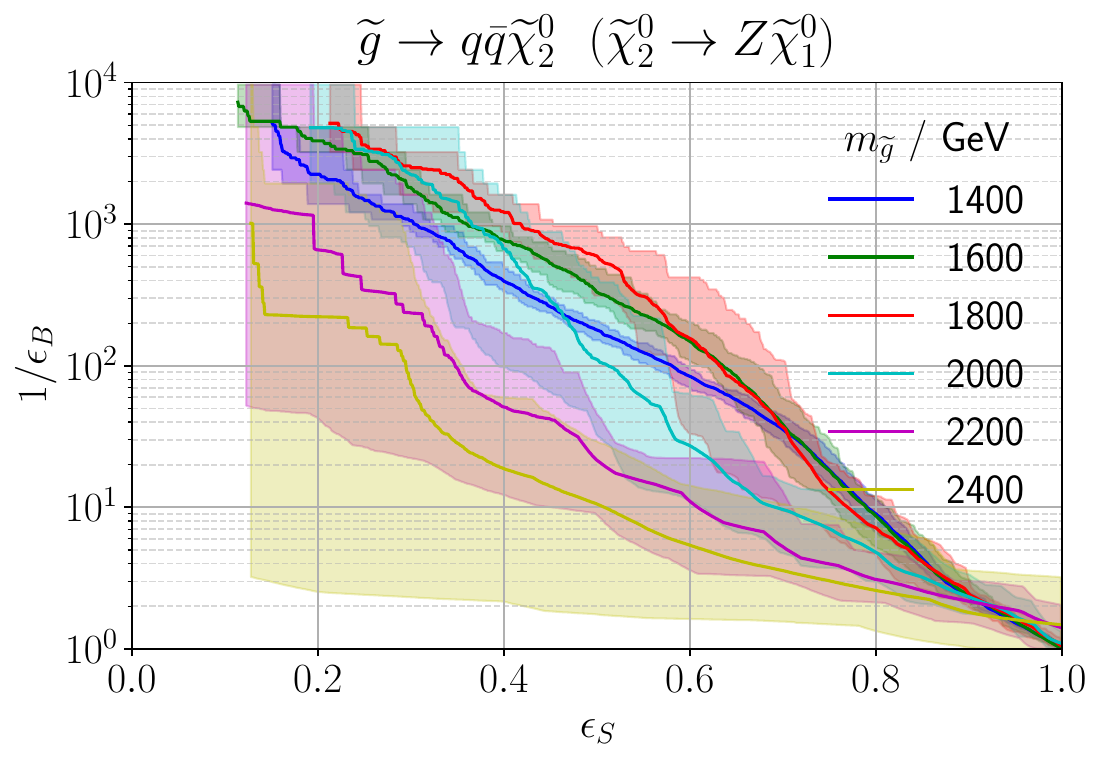}
\includegraphics[width=0.49\linewidth]{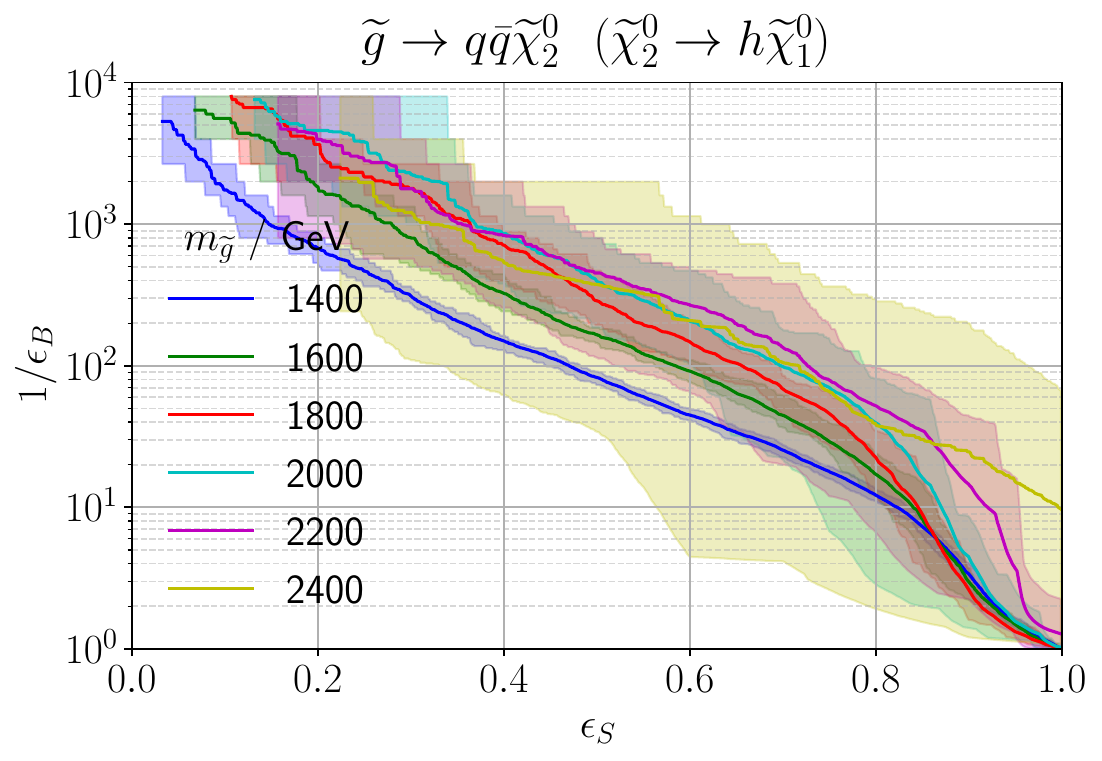}

\includegraphics[width=0.49\linewidth]{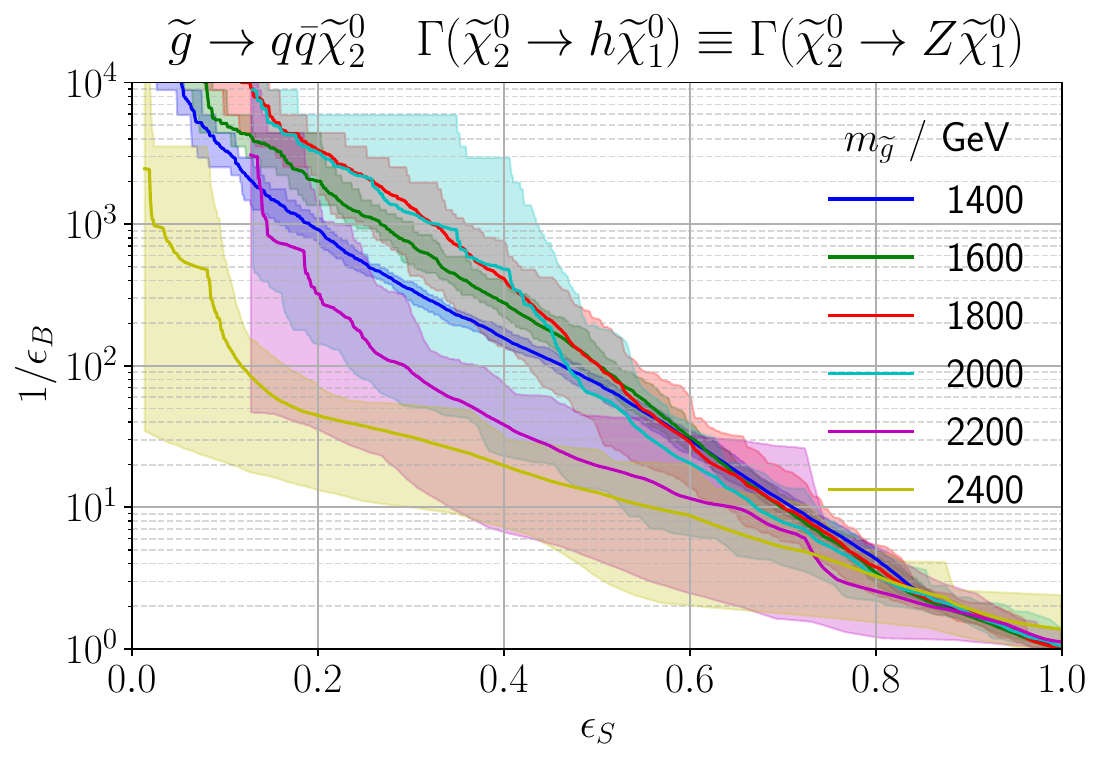}
\includegraphics[width=0.49\linewidth]{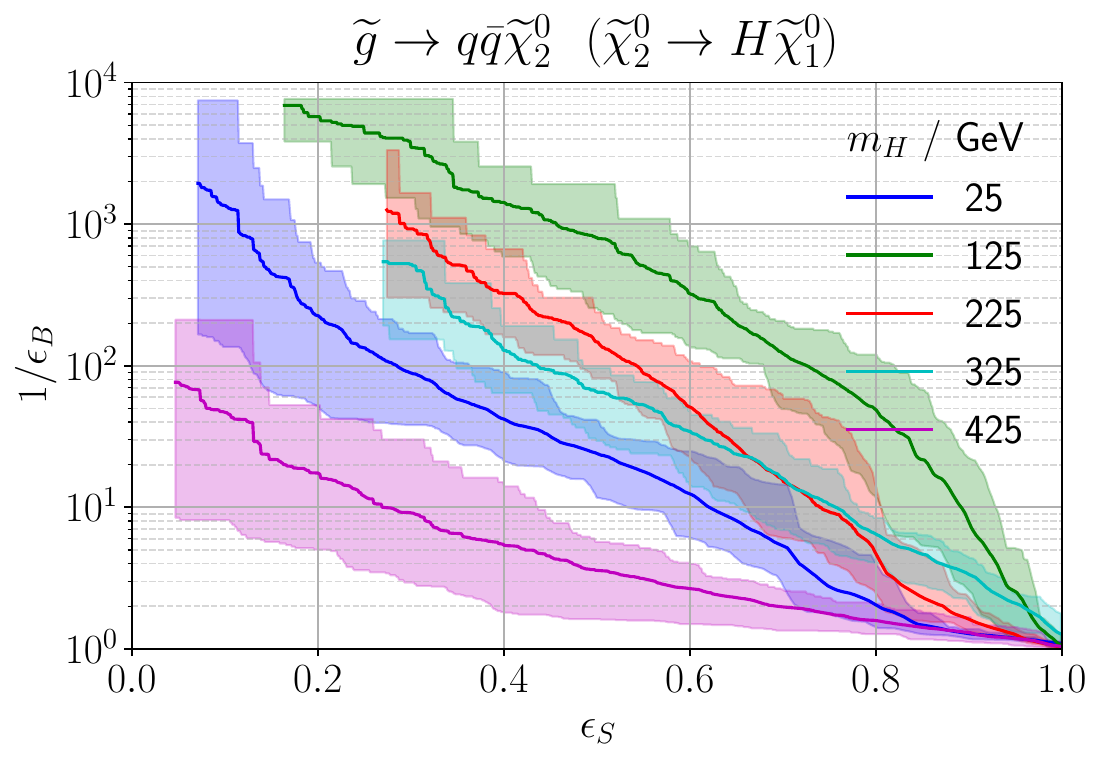}
\caption{ROC curves of all four benchmark models. Solid lines denote the mean value and shaded regions show the span between the minimum and maximum values obtained from ten different signal injections}
\label{fig:rocs}
\end{figure*}

In figure \ref{fig:rocs} we show ROC curves, i.e. background suppression as a function of signal efficiency of our benchmark models. We also observe a common feature of anomaly detection techniques. With rising signal cross section the classifier learns to better separate background from signal-like events. At the same time, larger signal cross sections correspond to smaller gluino masses, which in turn leads to less expressive features. Both effects combined lead to intermediate gluino masses having the largest background suppression at the same signal efficiency compared to small masses with large cross sections or large masses with very obvious signatures, especially in the decay to $Z$ and Standard Model Higgs bosons. We also observe in the bottom right figure that for low and high Higgs masses the background rejection is noticeably weaker than for intermediate masses. For light Higgs masses, the jets are too similar to background jets while high Higgs masses lead to wide jets that get reconstructed incorrectly.
\label{sec:roc_curve}

\bibliographystyle{unsrturl}
\bibliography{bibliography}
\end{document}